\documentclass[aps,pre,twocolumn,superscriptaddress]{revtex4-2}

\usepackage{graphicx}
\usepackage{amsmath}
\usepackage{mathrsfs}
\usepackage{indentfirst}
\usepackage{bm}
\usepackage{csquotes}
\begin{document}

\title{Atomic oven with rapid thermal response for atom experiments}

\author{Weilong Huang}
\author{Congjun Zou}
\author{Feiyu Dong}
\author{Huirong Xiao}
\affiliation {Key Laboratory of Atomic and Subatomic Structure and Quantum Control (Ministry of Education), Guangdong Basic Research Center of Excellence for Structure and Fundamental Interactions of Matter, and School of Physics, South China Normal University, Guangzhou 510006, China}
\affiliation {Guangdong Provincial Key Laboratory of Quantum Engineering and Quantum Materials, Guangdong-Hong Kong Joint Laboratory of Quantum Matter, and Frontier Research Institute for Physics, South China Normal University, Guangzhou 510006, China}

\author{Zejian Ren}
\affiliation{Quantum Science Center of Guangdong-Hong Kong-Macao Greater Bay Area, Shenzhen, China}
\author{Shanchao Zhang}\email[Email:]{sczhang@m.scnu.edu.cn}
\affiliation {Key Laboratory of Atomic and Subatomic Structure and Quantum Control (Ministry of Education), Guangdong Basic Research Center of Excellence for Structure and Fundamental Interactions of Matter, and School of Physics, South China Normal University, Guangzhou 510006, China}
\affiliation {Guangdong Provincial Key Laboratory of Quantum Engineering and Quantum Materials, Guangdong-Hong Kong Joint Laboratory of Quantum Matter, and Frontier Research Institute for Physics, South China Normal University, Guangzhou 510006, China}
\affiliation{Quantum Science Center of Guangdong-Hong Kong-Macao Greater Bay Area, Shenzhen, China}
\affiliation{GPETR Center for Quantum Precision Measurement, South China Normal University, Guangzhou 510006, China}

\date{\today}
\begin{abstract}
\indent Atomic oven generating controllable atomic beam flux plays a fundamental role in quantum gas experiments. Here, we report a new heater design that can heat up an high temperature atomic oven with fast thermal response. The new heater shows a heating rate improved by 7.65 times comparing to that of the conventional resistive heater while the crucible temperature can heated up to 1200K. With this oven, we generated a collimated ytterbium beam with flux exceeding $10^{14} \text{ atoms/s}$ at 823 K. We believe that our design offers a promising solution for shortening experimental dead time and improve the experiment efficiency in cold atom researches.
\end{abstract}
\maketitle

\section{Introduction}
\indent Neutral atom quantum gases constitute a central experimental platform for quantum many-body physics \textsuperscript{\cite{Bluvstein,Honda,Lee2024,Zhao,Zhou,Asteria,Koepsell,Meng}}, precision metrology \textsuperscript{\cite{Zaporski,Peng}}, and quantum computation \textsuperscript{\cite{Kiefer,Bojovic,Bluvstein2026}}. In a typical cold atom experiment, the preparation cycle begins with the generation of a high-flux atomic beam, followed by laser cooling and trapping, and evaporative cooling towards quantum degeneracy \textsuperscript{\cite{Lee2017,Stellmer,Ciamei,Wilson,Ye}}. While substantial efforts have been devoted to exploring cooling and trapping technique, progress of the atomic oven technique is still limited.

Conventionally, an atomic crucible in an effusion atomic oven is heated up by a resistive heater
 \textsuperscript{\cite{Song,Stan,Schioppo,Li,Vishwakarma,Ravensbergen,Hirzler}} via the heat conduction in ultrahigh vacuum, which is surrounded by complicated heat insulation structure. A typical precedure of heating up or cooling down the atomic oven may take as long as hours.  This slow heat response usually leads to prolonged apparatus preparation time in table-top cold atom experiments like ytterbium (Yb), erbium(Er) and dysprosium (Dy) with oven temperature ranging from 700~K and to 1200~K \textsuperscript{\cite{Fukuhara,Lu2012}}.


\indent In this work, we demonstrate a compact atomic oven based on inductive heating with fast thermal response. Our design utilizes electromagnetic waves to generate eddy current in the wall of atomic crucible and thus only heat up the crucible itself directly. We characterize the source performance by generation Yb atomic beam flux, which shows a 7.65-fold improvement in heating rate and a stable flux of $10^{14} \text{ atoms/s}$. A physical model is developed to describe the power transfer efficiency and skin-depth scaling, confirming that inductive heating significantly enhances the dynamic range of thermal control for cold atom experiments.

\section{design and constrution}
\indent The atomic oven consists of three main components: a stainless-steel atomic crucible, a capillary array collimating nozzle, and an inductive heating system. As shown in the inset of Fig. 1, the crucible is machined from ultrahigh vacuum compatible 304 stainless steel and mounted on a flange. Its cylindrical geometry is designed by balancing radio-frequency skin depth and vacuum machining constraints.

\begin{figure}[h]
    \centering
    \includegraphics[width=0.47\textwidth]{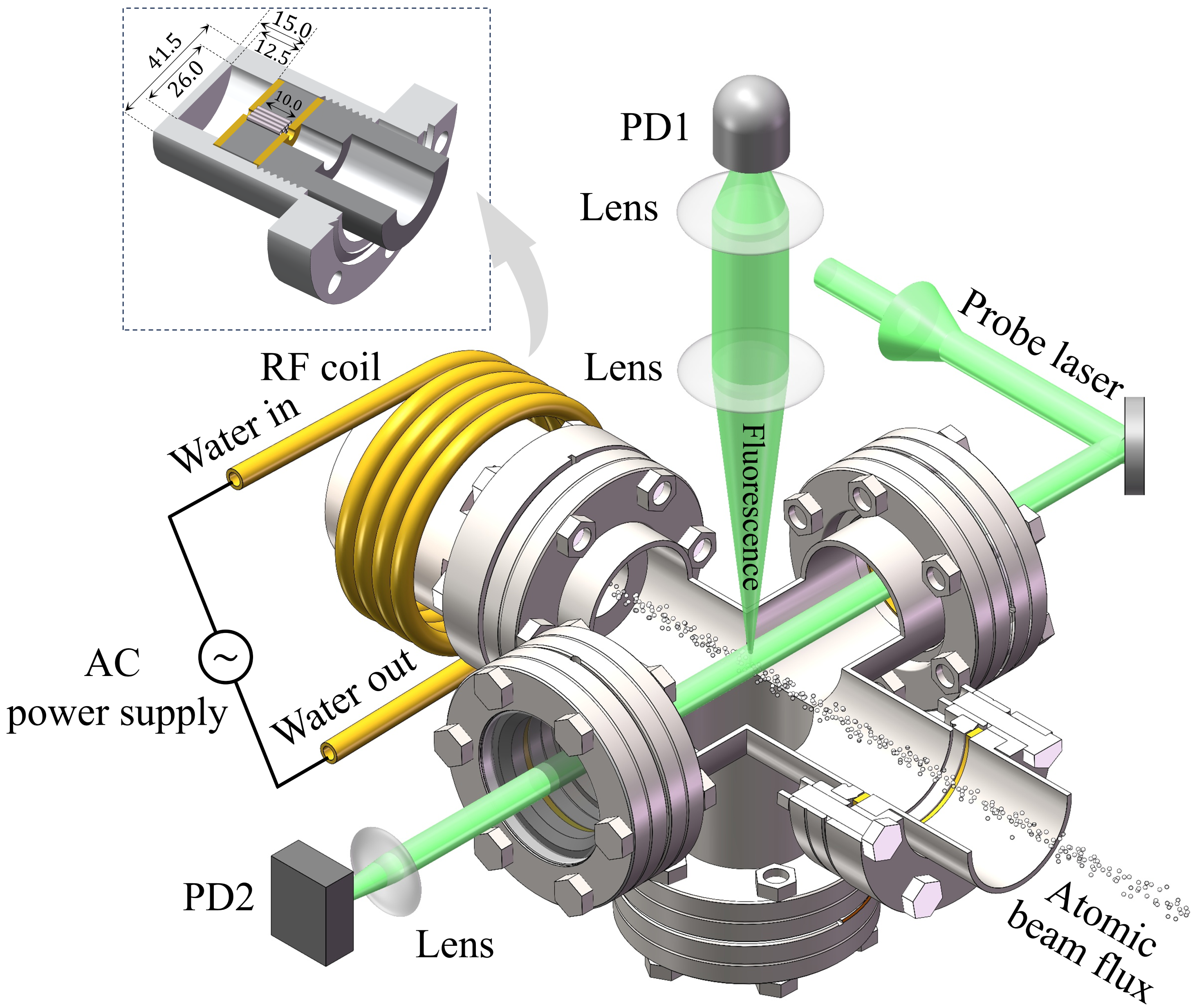}
    \caption{Experimental setup of the inductively heated atomic source and optical detection system. A radio-frequency-driven copper induction coil surrounds the atomic crucible, while fluorescence and absorption of the atomic beam are probed downstream using a resonant laser. Inset: Schematic of the atomic crucible and collimating nozzle geometry. The crucible is machined from 304 stainless steel and optimized by balancing eddy current penetrating depth considerations and vacuum machining constraints. Key geometrical parameters, including crucible wall thickness and capillary dimensions, are indicated in units of millimeters (mm). }
\end{figure}

The collimating nozzle employs a hexagonally packed array of stainless-steel capillaries with inner diameter of 0.2~mm, outer diameter of 0.4~mm, and length of 10~mm, resulting in a collimation angle of approximately 20~mrad. This configuration suppresses stray atomic deposition and improves beam collimation \textsuperscript{\cite{Hanes,Senaratne}}.

\indent Inductive heating \textsuperscript{\cite{Lu2019,Lu2021}} is realized by driving a water-cooled hollowed copper coil with alternating current. The coil has a winding diameter of 65~mm and an inter-turn spacing of 2~mm, as shown in Fig. 1. Temperature is monitored using a type-K thermocouple and stabilized by a proportional-integral (PI) feedback servo. A single-layer ceramic housing provides thermal insulation while maintaining a compact structure.

\indent To understand the enhanced thermal response, we model the power injection into the crucible. The inductive heating efficiency is governed by the eddy current at the outer surface of metalic crucible\textsuperscript{\cite{Jackson,Rudnev}}, where the penetrating depth $\delta$ is given by:
\begin{equation}
\delta = \sqrt{\frac{2\rho}{\omega \mu}}
 \end{equation}

 where $\rho$ is the electrical resistivity of the crucible material, $\omega$ is the angular frequency of the driving current, and $\mu$ is the magnetic permeability. For our operating frequency range, the penetrating depth is matched to the crucible wall thickness to maximize the heat power conducted to the crucible wall $P_{eddy}$.

\indent The temperature change of the crucible is described by the below equation:

\begin{equation}
C_p m \frac{dT}{dt} = P_{in} - P_{loss}
\end{equation}
	where $C_p$ is the specific heat capacity and $P_{loss}$ represents radiative and conductive losses. In resistive systems, $P_{in}$ is limited by the thermal contact conductance between the filament and the crucible. In contrast, our inductive design generates heat in crucible directly and allowing $P_{in}$ to be limited only by the radio-frequency power coupling efficiency, resulting in the significantly larger temperature gradient $dT/dt$ observed in our results.

\indent The atomic crucible has a cylindrical geometry with base diameter of 26 mm and internal volume of $6.7 \text{ cm}^3$ machined from ultrahigh vacuum compatible 304 stainless steel. To minimize conductive heat loss while ensuring structural integrity, we customized a ceramic housing with inner diameter of 45.5 mm as the crucible holder and also thermal break. Ceramic material is selected for its high electrical resistivity as an insulator and moderate thermal conductivity, which provides a balance between insulation and preventing thermal runaway during cooling phases.
 \begin{figure}[hb]
	\centering
	\includegraphics[width=0.47\textwidth]{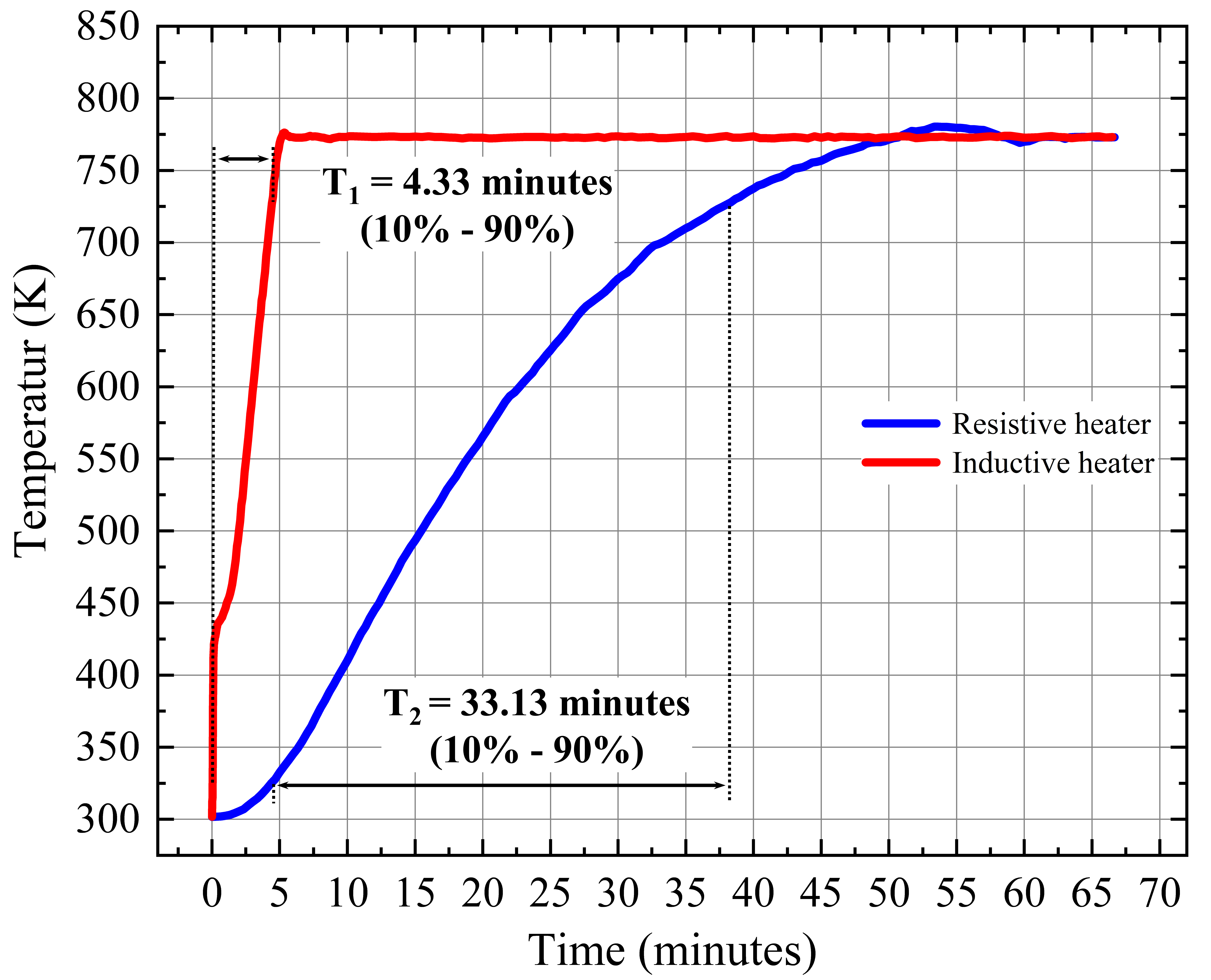}
	\caption{Time-resolved temperature evolution of the atomic crucible under resistive heater (blue) and inductive heater (red), measured under identical geometrical configurations and target temperature (773 K). The time constants $T_1$ and $T_2$ represent the $10\,\%$--$90\,\%$ rise times, indicating the significantly faster thermal response of the inductive heating method.}
\end{figure}

\section{characterization and results}
\indent We performed a comparative study between inductive and resistive heating under identical insulation and geometrical conditions. The target temperature was set to 773 K. As shown in Fig. 2, our inductive heater achieved the target temperature in 4.33 minutes, comparing to 33.13 minutes for the resistive heater. The initial heating rate $dT/dt$ for the inductive heater was improved by approximately 7.65 times. This significant thermal response manifests as the inductive heater maintaining a substantially higher heating rate than the resistive heater throughout the process. 

Upon initial heating, the low electrical resistivity compresses the skin depth to a minimum, confining the induced currents tightly to the surface and establishing an extreme energy density, resulting in a drastically steeper initial temperature rise. As the temperature increases, the rising resistance change eddy current profile inside the crucible, lowering the volumetric energy density and reducing the total heating rate. In contrast, resistive heater relies on bulk thermal conduction with slow heat accumulation from the surface inward, leaving it persistently lower than that of inductive heating. Meanwhile, the resitive heater shows clear overshoot before the temperature is finally stablized while the inductive heater has a ignorable overshoot due to its fast thermal response.

\begin{figure}[ht]
    \centering
    \includegraphics[width=0.43\textwidth]{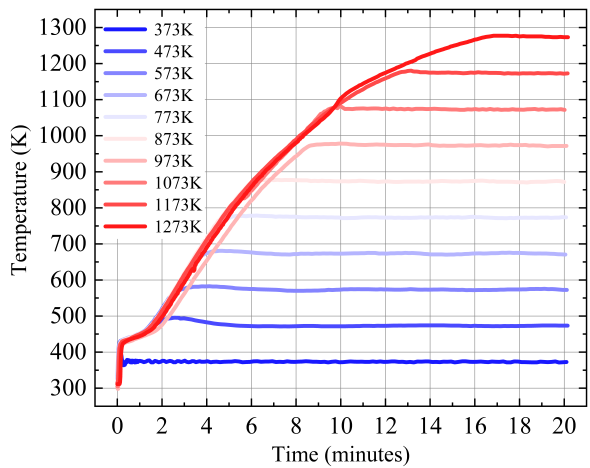}
    \caption{Wide-range thermal setpoint-programmed heating. Time-resolved temperature profiles demonstrating the system's ability to rapidly reach and maintain precise target temperatures across an extensive range from $373\,\mathrm{K}$ to $1273\,\mathrm{K}$. This confirms not only the fast thermal response but also the high maximum operating temperature and excellent long-term stability of the inductive heating setup.}
\end{figure}

We also characterize the heating performance by measuring the temperature curve during heating up procedure with different targeted temperature, as shown in Fig. 3. Our inductive heating oven performs robust in this repeatedly test.  Temperature stability is also crucial for an operating atomic oven. With our inductive heater, the stablized temperature shows a fluctuation of $\Delta T < \pm 2 \text{ K}$ at 1273 K, which is achieved by using a PID feedback loop working in the pulse width modulation mode that swith on and off the driving radio-frequency in the coil. The stability of this rapid thermal response is maintained across the entire operational range, demonstrating the robustness and feasibility of the control scheme from low temperatures to high temperatures.

\indent Eventually, we utilized absorption spectroscopy of Yb at its ${}^{1}S_{0} \rightarrow {}^{3}P_{1}$ transition to characterize the generated atomic beam flux. Fig. 4(a) presents the fluorescence spectroscopy used to identify isotope shifts. The transverse velocity derived from the linewidth is approximately $\pm$72 m/s, indicating a well-collimated, low-divergence atomic beam that enhances spectral resolution and minimizes Doppler broadening. Quantitative flux measurements were derived from the transmission signal. To quantify each Yb isotope's flux, we use the strongest $^{174}$Yb resonance fluorescence peak in Fig. 4(a) as a reference, since its intensity scales with beam density. Other isotopes' fluxes are then determined by comparing their peak intensities to this dominant peak. Fig. 4(b) displays the measured flux of isotope $^{174}\mathrm{Yb}$ as a function of crucible temperature. The data follow the expected exponential dependence derived from the Arrhenius equation for vapor pressure, confirming the thermodynamic consistency of the source operation. At a temperature of 823 K, we measured a flux exceeding $1.0 \times 10^{14} \text{ atoms/s}$. This flux level is sufficient to enable efficient loading of a magneto-optical trap, making the source suitable for ultracold atom experiments.

  \begin{figure}[ht]
    \centering
    \includegraphics[width=0.43\textwidth]{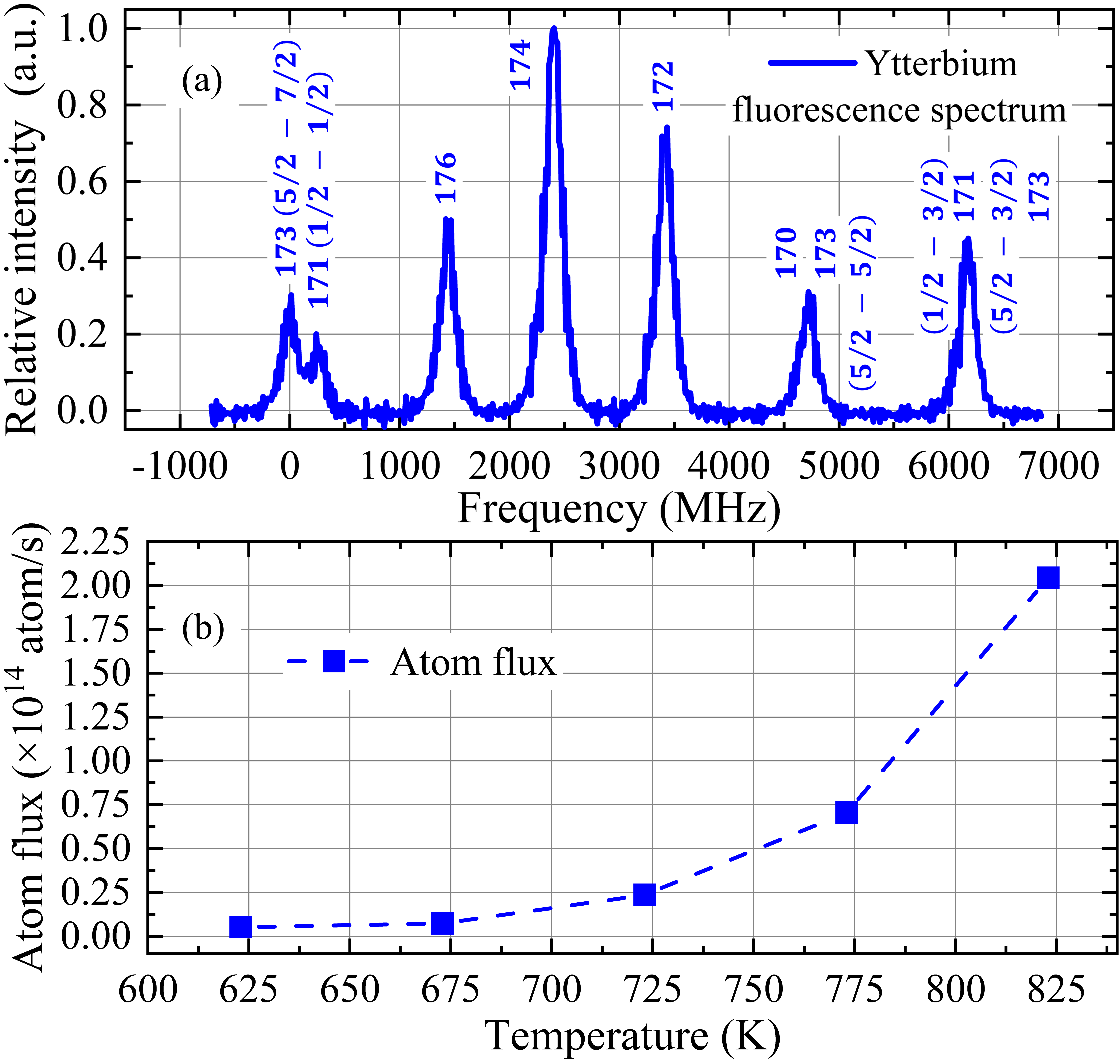}
    \caption{Characterization of the atomic source. (a) Resonance fluorescence spectra of Yb isotopes. The probe laser is scanned across the isotopic transitions, and the fluorescence is collected by  Photodetector1(PD1) employing the fluorescence collection optics depicted in Fig. 1. (b) Atomic beam flux versus heating temperature. The dashed line is a guide to the eye. The probe laser propagates perpendicularly through the atomic beam, and the transmitted intensity monitored by PD2 is used to derive the atomic beam density.}
\end{figure}

\section{Concluding remarks}
\indent We have demonstrated an inductively heated atomic source that satisfies the stringent requirements of modern AMO experiments. By achieving a 7.65-fold increase in heating rate and maintaining high thermal stability, this design significantly reduces experimental dead time. The theoretical analysis confirms that inductive coupling offers a superior scaling for power transfer compared to contact-based resistive heating for refractory elements. This compact, fast-response source is well-suited for the next generation of high-repetition-rate quantum gas microscopes and optical lattice clocks.

\section{Data availability}
\indent The data that support the findings of this study are available from the corresponding author upon reasonable request.

{\bf Acknowledgement} This work is supported by the National Key R\&D Program of China (2022YFA1405300); Guangdong Provincial Quantum Science Strategic Initiative (GDZX2404001, GDZX2504005); National Natural Science Foundation of China (12322408); Guangdong Basic and Applied Basic Research Foundation (2026A1515030019, 2025A1515011684).


\begin{thebibliography}{}
\bibitem{Bluvstein} D. Bluvstein, A. Omran, H. Levine, A. Keesling, G. Semeghini, S. Ebadi, T. T. Wang, A. A. Michailidis, N. Maskara, W. W. Ho, S. Choi, M. Serbyn, M. Greiner, V. Vuletic, and M. D. Lukin, Controlling quantum many-body dynamics in driven Rydberg atom arrays, Science \textbf{371}, 1355-1359 (2021).

\bibitem{Asteria} L. Asteria, H. P. Zahn, M. N. Kosch, K. Sengstock, and C. Weitenberg, Quantum gas magnifier for sub-lattice-resolved imaging of 3D quantum systems, Nature \textbf{599}, 571-575 (2021).

\bibitem{Koepsell} J. Koepsell, D. Bourgund, P. Sompet, S. Hirthe, A. Bohrdt, Y. Wang, F. Grusdt, E. Demler, G. Salomon, C. Gross, and I. Bloch, Microscopic evolution of doped Mott insulators from polaronic metal to Fermi liquid, Science \textbf{374}, 82-86 (2021).

\bibitem{Zhou} T.-W. Zhou, G. Cappellini, D. Tusi, L. Franchi, J. Parravicini, C. Repellin, S. Greschner, M. Inguscio, T. Giamarchi, M. Filippone, J. Catani, and L. Fallani, Observation of universal Hall response in strongly interacting Fermions, Science \textbf{381}, 427-430 (2023).

\bibitem{Meng} Z. Meng, L. Wang, W. Han, F. Liu, K. Wen, C. Gao, P. Wang, C. Chin, and J. Zhang, Atomic Bose-Einstein condensate in twisted-bilayer optical lattices, Nature \textbf{615}, 231-236 (2023).

\bibitem{Lee2024} K. Lee, S. Kim, T. Kim, and Y. Shin, Universal Kibble-Zurek scaling in an atomic Fermi superfluid, Nat. Phys. \textbf{20}, 1570-1574 (2024).

\bibitem{Honda} K. Honda, Y. Takasu, S. Goto, H. Kazuta, M. Kunimi, I. Danshita, and Y. Takahashi, Observation of slow relaxation due to Hilbert space fragmentation in strongly interacting Bose-Hubbard chains, Sci. Adv. \textbf{11}, eadv3255 (2025).

\bibitem{Zhao} E. Zhao, Z. Wang, C. He, T. F. J. Poon, K. K. Pak, Y.-J. Liu, P. Ren, X.-J. Liu, and G.-B. Jo, Two-dimensional non-Hermitian skin effect in an ultracold Fermi gas, Nature \textbf{637}, 565-573 (2025).

\bibitem{Zaporski} L. Zaporski, Q. Liu, G. Velez, M. Radzihovsky, Z. Li, S. Colombo, E. Pedrozo-Penafiel, and V. Vuletic, Quantum-amplified global-phase spectroscopy on an optical clock transition, Nature \textbf{646}, 309-314 (2025).

\bibitem{Peng} S. Peng, S. Peng, L. Ren, S. Liu, B. Liu, J. Li, and L. Luo, Precision Measurement of Spin-Dependent Dipolar Splitting in $^6$Li p-Wave Feshbach Resonances, Phys. Rev. Lett. \textbf{135}, 133401 (2025).

\bibitem{Kiefer} Y. Kiefer, Z. Zhu, L. Fischer, S. Jele, M. Gachter, G. Bisson, K. Viebahn, and T. Esslinger, Protected quantum gates using qubit doublons in dynamical optical lattices, Nature (2026).

\bibitem{Bojovic} P. Bojovic, T. Hilker, S. Wang, J. Obermeyer, M. Barendregt, D. Tell, T. Chalopin, P. M. Preiss, I. Bloch, and T. Franz, High-fidelity collisional quantum gates with fermionic atoms, Nature (2026).

\bibitem{Bluvstein2026} D. Bluvstein, A. A. Geim, S. H. Li, S. J. Evered, J. P. Bonilla Atiades, G. Baranes, A. Gu, T. Manovitz, M. Xu, M. Kalinowski, S. Majidy, C. Kokail, N. Maskara, E. C. Trapp, L. M. Stewart, S. Hollerith, H. Zhou, M. J. Gullans, S. F. Yelin, M. Greiner, V. Vuletic, M. Cain, and M. D. Lukin, A fault-tolerant neutral-atom architecture for universal quantum computation, Nature \textbf{649}, 39-46 (2026).

\bibitem{Stellmer} S. Stellmer, R. Grimm, and F. Schreck, Production of quantum-degenerate strontium gases, Phys. Rev. A \textbf{87}, 013611 (2013).

\bibitem{Lee2017} M. Lee, J. H. Han, J. H. Kang, M.-S. Kim, and Y. Shin, Double resonance of Raman transitions in a degenerate Fermi gas, Phys. Rev. A \textbf{95}, 043627 (2017).

\bibitem{Ye} Z.-X. Ye, L.-Y. Xie, Z. Guo, X.-B. Ma, G.-R. Wang, L. You, and M. K. Tey, Double-degenerate Bose-Fermi mixture of strontium and lithium, Phys. Rev. A \textbf{102}, 033307 (2020).

\bibitem{Wilson} K. E. Wilson, A. Guttridge, J. Segal, and S. L. Cornish, Quantum degenerate mixtures of Cs and Yb, Phys. Rev. A \textbf{103}, 033306 (2021).

\bibitem{Ciamei} A. Ciamei, S. Finelli, A. Cosco, M. Inguscio, A. Trenkwalder, and M. Zaccanti, Double-degenerate Fermi mixtures of $^6$Li and $^{53}$Cr atoms, Phys. Rev. A \textbf{106}, 053318 (2022).

\bibitem{Stan} C. A. Stan and W. Ketterle, Multiple species atom source for laser-cooling experiments, Rev. Sci. Instrum. \textbf{76}, 063113 (2005).

\bibitem{Schioppo} M. Schioppo, N. Poli, M. Prevedelli, St. Falke, Ch. Lisdat, U. Sterr, and G. M. Tino, A compact and efficient strontium oven for laser-cooling experiments, Rev. Sci. Instrum. \textbf{83}, 103101 (2012).

\bibitem{Song} B. Song, Y. Zou, S. Zhang, C.-w. Cho, and G.-B. Jo, A cost-effective high-flux source of cold ytterbium atoms, Appl. Phys. B \textbf{122}, 250 (2016).

\bibitem{Ravensbergen} C. Ravensbergen, V. Corre, E. Soave, M. Kreyer, and R. Grimm, Production of a degenerate Fermi-Fermi mixture of dysprosium and potassium atoms, Phys. Rev. A \textbf{98}, 063624 (2018).

\bibitem{Vishwakarma} C. Vishwakarma, J. Mangaonkar, K. Patel, G. Verma, S. Sarkar, and U. D. Rapol, A simple atomic beam oven with a metal thermal break, Rev. Sci. Instrum. \textbf{90}, 053106 (2019).

\bibitem{Hirzler} H. Hirzler, T. Feldker, H. Furst, N. V. Ewald, E. Trimby, R. S. Lous, J. D. Arias Espinoza, M. Mazzanti, J. Joger, and R. Gerritsma, Experimental setup for studying an ultracold mixture of trapped Yb$^+$-$^6$Li, Phys. Rev. A \textbf{102}, 033109 (2020).

\bibitem{Li} J. Li, Z.-P. Jia, P. Liu, X.-Y. Liu, D.-Z. Wang, D.-Q. Kong, S.-P. Li, X.-Y. Cui, H.-N. Dai, Y.-A. Chen, and J.-W. Pan, An integrated high-flux cold atomic beam source for strontium, Rev. Sci. Instrum. \textbf{94}, 093202 (2023).

\bibitem{Fukuhara} T. Fukuhara, Y. Takasu, M. Kumakura, and Y. Takahashi, Degenerate Fermi Gases of Ytterbium, Phys. Rev. Lett. \textbf{98}, 030401 (2007).

\bibitem{Lu2012} M. Lu, N. Q. Burdick, and B. L. Lev, Quantum Degenerate Dipolar Fermi Gas, Phys. Rev. Lett. \textbf{108}, 215301 (2012).

\bibitem{Hanes} G. R. Hanes, Multiple tube collimator for gas beams, J. Appl. Phys. \textbf{31}, 2171-2175 (1960).

\bibitem{Senaratne} R. Senaratne, S. V. Rajagopal, Z. A. Geiger, K. M. Fujiwara, V. Lebedev, and D. M. Weld, Effusive atomic oven nozzle design using an aligned microcapillary array, Rev. Sci. Instrum. \textbf{86}, 023105 (2015).

\bibitem{Lu2019} W. Lu, L. T. Sun, C. Qian, L. B. Li, J. W. Guo, W. Huang, X. Z. Zhang, and H. W. Zhao, Production of intense uranium beams with inductive heating oven at Institute of Modern Physics, Rev. Sci. Instrum. \textbf{90}, 113318 (2019).

\bibitem{Lu2021} W. Lu, C. Qian, W. H. Zhang, H. Y. Ma, J. D. Ma, Y. C. Feng, L. B. Li, L. X. Li, J. W. Guo, W. Huang, X. Z. Zhang, L. T. Sun, and H. W. Zhao, Production of metallic ion beams by electron cyclotron resonance ion sources equipped with inductive heating ovens at the Institute of Modern Physics, Rev. Sci. Instrum. \textbf{92}, 033302 (2021).

\bibitem{Jackson} J. D. Jackson, \textit{Classical Electrodynamics}, 3rd ed. (Wiley, New York, 1999).
	
\bibitem{Rudnev} V. Rudnev, D. Loveless, R. Cook, and M. Black, \textit{Handbook of Induction Heating} (CRC Press, 2003).

\end{thebibliography}
\end{document}